# SEPARATION OF MULTIPLE EVOKED RESPONSES USING DIFFERENTIAL AMPLITUDE AND LATENCY VARIABILITY

*Kevin H. Knuth*[1], *Wilson A. Truccolo*[2], *Steven L. Bressler*[2], *Mingzhou Ding*[2]

[1] Center for Advanced Brain Imaging and Cognitive Neuroscience and Schizophrenia Dept.,
Nathan S. Kline Institute, Orangeburg NY 10962
[2] Center for Complex Systems and Brain Sciences, Florida Atlantic University, Boca Raton FL 33431

**ABSTRACT**

In neuroelectrophysiology one records electric potentials or magnetic fields generated by ensembles of synchronously active neurons in response to externally presented stimuli. These evoked responses are often produced by multiple generators in the presence of ongoing background activity. While source localization techniques or current source density estimation are usually used to identify generators, application of blind source separation techniques to obtain independent components has become more popular.

We approach this problem by applying the Bayesian methodology to a more physiologically-realistic source model. As it is generally accepted that single trials vary in amplitude and latency, we incorporate this variability into the model. Rather than making the unrealistic assumption that these cortical components are independent of one another, our algorithm utilizes the differential amplitude and latency variability of the evoked waveforms to identify the cortical components. The algorithm is applied to intracortically-recorded local field potentials in monkeys performing a visuomotor task.

## 1. INTRODUCTION

The techniques of neuroelectrophysiology rely on the recording of electric potentials or magnetic fields evoked in the brain during the presentation of stimuli in cognitive or sensorimotor tasks. These evoked responses are often generated by multiple ensembles of neurons firing synchronously in response to the presented stimuli. Far from being independent, these neural ensembles, also referred to as generators or sources, are often dynamically coupled in unknown ways that are of interest to the experimenter. Thus the recording channels, electrodes in electroencephalography (EEG) and superconducting quantum interference devices (SQUIDs) in magnetoencephalography (MEG), record linear mixtures of evoked responses from these sources approximately time-locked to the experimental stimulus in addition to ongoing background activity. Because the mixing is linear, instantaneous and often stationary, recent developments in linear blind source separation (BSS) and independent component analysis (ICA) have been useful in analyzing EEG and MEG signals using both ensemble averaged data [1,2,3] and single-trials [4,5], when applied with care [6].

However, by assuming independence of the sources, the experimenter assumes away one of the most interesting aspects of the active neural ensembles in the brain, the nature of their dynamical interactions. In addition, by working with ensemble averages of the responses, which is typically done to improve the signal to noise ratio, one is implicitly assuming that the evoked waveform is identical in all respects in every experimental trial. In this work we introduce a more realistic model of the evoked response, one which includes the possibility of trial-to-trial variability of source amplitude and latency. By adopting this modeling approach, we find that we can (i) more accurately account for trial-to-trial variability [7], (ii) utilize the differential variation in the amplitudes and latencies to identify sources, (iii) avoid enforcing statistical independence of the sources, and (iv) more accurately estimate the ongoing background activity.

## 2. MODELING EVOKED RESPONSES

We model the evoked response from a single source by assuming that the signal has a stereotypic waveshape, but can vary in amplitude from trial-to-trial. In addition, the response is not assumed to be strictly time-locked to the stimulus, rather the onset latency of the response can vary from trial-to-trial. We write this mathematically as

Thanks to the Charles E. Schmidt College of Science computing facilities at FAU. Supported by NIMH, NSF, ONR and a CNPq (Brazil) fellowship.

$\alpha s(t-\tau)$, where $s(\cdot)$ represents the stereotypic waveshape of the response, $\alpha$ represents the amplitude of the response in that trial and $\tau$ represents the onset latency shift. As we are estimating both the waveshape and the amplitude and latency, there is a degeneracy in the model. To eliminate this degeneracy, we take as a convention that the ensemble average amplitude of the response over the recorded trials is unity and the ensemble average latency is zero.

As described above there may be multiple neural sources, as well as multiple detectors. To describe the source-detector coupling we introduce a coupling matrix, $\mathbf{C}$, which is commonly known as a mixing matrix in BSS. In addition, an experimenter records many trials. For the $r^{th}$ recorded trial we write the signal recorded in the $m^{th}$ detector in component form as

$$x_{mr}(t) = \sum_{n=1}^{N} C_{mn} \alpha_{nr} s_n(t - \tau_{nr}) + \eta_{mr}(t), \quad (1)$$

where $n$ indexes the $N$ neural sources, $C_{mn}$ is the coupling between the $m^{th}$ detector and the $n^{th}$ source, $\alpha_{nr}$ is the amplitude of the $n^{th}$ source during the $r^{th}$ trial, $\tau_{nr}$ is the latency of the $n^{th}$ source during the $r^{th}$ trial, $s_n(\cdot)$ is the waveshape of the $n^{th}$ source, and $\eta_{mr}(t)$ is the unpredictable signal component recorded in the $m^{th}$ detector during the $r^{th}$ trial. This unpredictable signal component is a combination of the recorded background activity along with any noise in the channel. For simplicity, we assume that this unpredictable signal component has zero mean.

## 3. BAYESIAN DERIVATION OF THE ALGORITHM

Bayes' Theorem is the natural starting point because it allows one to describe the probability of the model in terms of the likelihood of the data and the prior probability of the model

$$p(model \mid data, I) = \frac{p(data \mid model, I) \, p(model \mid I)}{p(data \mid I)}, \quad (1)$$

where $I$ represents any prior information one may have about the physical situation. Bayes' Theorem can be viewed as describing how one's prior probability, $P(model \mid I)$, is modified by the acquisition of some new information.

To apply this to our problem, we consider the change in our knowledge about the model with the acquisition of new data consisting of a set of recorded trials $\mathbf{x}(t)$ recorded by a set of detectors. In this case, Bayes' Theorem can be written as

$$p(\mathbf{C}, \mathbf{s}(t), \boldsymbol{\alpha}, \boldsymbol{\tau} \mid \mathbf{x}(t), I) = \quad (2)$$

$$\frac{p(\mathbf{x}(t) \mid \mathbf{C}, \mathbf{s}(t), \boldsymbol{\alpha}, \boldsymbol{\tau}, I) \, p(\mathbf{C}, \mathbf{s}(t), \boldsymbol{\alpha}, \boldsymbol{\tau} \mid I)}{p(\mathbf{x}(t) \mid I)},$$

where boldface symbols represent the entire set of parameters of each type, eg. $\boldsymbol{\alpha} = \{\alpha_1, \alpha_2, ..., \alpha_R\}$. As we would like to find the model that maximizes the probability in Equation (2), in practice we rewrite the equation as a proportionality and equate the inverse of the prior probability of the data $p(\mathbf{x}(t) \mid I)$ to the implicit proportionality constant

$$p(\mathbf{C}, \mathbf{s}(t), \boldsymbol{\alpha}, \boldsymbol{\tau} \mid \mathbf{x}(t), I) \propto \quad (3)$$

$$p(\mathbf{x}(t) \mid \mathbf{C}, \mathbf{s}(t), \boldsymbol{\alpha}, \boldsymbol{\tau}, I) \, p(\mathbf{C}, \mathbf{s}(t), \boldsymbol{\alpha}, \boldsymbol{\tau} \mid I).$$

The probability on the left-hand side of Equation (3) is referred to as the posterior probability. It represents the probability that a given set of hypothesized values of the model parameters accurately describes the physical situation. The first term in the on the right-hand side is the likelihood of the data given the model. It describes the degree of accuracy with which we believe the model can predict the data. The final term in the numerator is the joint prior probability of the model, also called the prior. This prior describes the degree to which we believe the model to be correct based only on our prior information about the problem. It is through the assignment of the likelihood and priors that we express all of our knowledge about the particular source separation problem.

For simplicity, the joint prior can be factored into four terms each representing independent physical processes,

$$p(\mathbf{C}, \mathbf{s}(t), \boldsymbol{\alpha}, \boldsymbol{\tau} \mid \mathbf{x}(t), I) \propto \quad (4)$$

$$p(\mathbf{x}(t) \mid \mathbf{C}, \mathbf{s}(t), \boldsymbol{\alpha}, \boldsymbol{\tau}, I) \, p(\mathbf{C} \mid I) \, p(\mathbf{s}(t) \mid I) \, p(\boldsymbol{\alpha} \mid I) \, p(\boldsymbol{\tau} \mid I).$$

For the amplitude and latency priors, we assign uniform densities with appropriate cutoffs denoting a range of physiologically realizable values.

At this point the relationship between our emerging algorithm (disregarding trial-to-trial variability) and the popular ICA algorithm introduced by Bell and Sejnowski [8] can be most easily noted. In the Bayesian derivation of ICA [6,9] one assigns a delta function likelihood expressing noise-free linear mixing, as well as a source amplitude prior (often a super- or sub-Gaussian probability density). With these two assignments, one can easily obtain a posterior probability for the mixing matrix alone by marginalizing over all possible source amplitudes. A gradient ascent method to find the most probable separation matrix completes the derivation.

Our derivation continues by utilizing the principle of maximum entropy to assign a Gaussian likelihood [10,11] by introducing a parameter $\sigma$ reflecting the expected square-deviation between our predictions and the mean

$$p(\mathbf{C}, \mathbf{s}(t), \boldsymbol{\alpha}, \boldsymbol{\tau}, \sigma \mid \mathbf{x}(t), I) \propto \quad (5)$$

$$\left(2\pi\sigma^2\right)^{\frac{-MRT}{2}} Exp\left[-\frac{1}{2\sigma^2}Q\right] p(\sigma \mid I) p(\mathbf{C} \mid I) p(\mathbf{s} \mid I),$$

where $p(\sigma \mid I)$ is the prior probability for $\sigma$ and $Q$ represents the sum of the square of the residuals between the data and our model in (1)

$$Q = \sum_{m=1}^{M}\sum_{r=1}^{R}\sum_{t=1}^{T}\left(x_{mr}(t) - \sum_{n=1}^{N} C_{mn}\alpha_{nr} s_n(t-\tau_{nr})\right)^2, \quad (6)$$

with $M$ representing the number of detectors, $R$ the number of experimental trials, and $T$ the number of recorded time points per trial. Assigning a Jeffreys prior for $\sigma$, $p(\sigma \mid I) = \sigma^{-1}$, and marginalizing the joint posterior over all possible values of $\sigma$ we obtain a marginal posterior for our original set of model parameters

$$p(\mathbf{C}, \mathbf{s}(t), \boldsymbol{\alpha}, \boldsymbol{\tau} \mid \mathbf{x}(t), I) \propto Q^{\frac{-MRT}{2}} p(\mathbf{C} \mid I) p(\mathbf{s} \mid I), \quad (7)$$

which is related to the Student t-distribution. Note that the uncertainty in our predictions expressed as $\sigma$ is not only dependent on the noise covariance, but is also dependent on the potential uncertainties in our measurements and the inadequacies of our model to describe the physical situation.

At this point prior information regarding the source waveforms could be used to further constrain the possible solutions. In addition, knowledge of the source-detector coupling, which is found by solving the electromagnetic forward problem, could be utilized to create an algorithm that simultaneously performs source separation and localization [12,13]. For simplicity, we choose rather to assume complete ignorance and assign uniform priors. The logarithm of the posterior probability can be compactly written as

$$\ln P = -\frac{MRT}{2}\ln Q + const, \quad (8)$$

where $P$ is the posterior probability $p(\mathbf{C}, \mathbf{s}(t), \boldsymbol{\alpha}, \boldsymbol{\tau} \mid \mathbf{x}(t), I)$.

The algorithm is completed by solving for the most probable set of model parameters, also called the *Maximum A Posteriori* (MAP) estimate. Examining the first partial derivative of the log posterior with respect to the $j^{th}$ source waveshape at time $q$ gives

$$\frac{\partial \ln P}{\partial s_j(q)} = -\frac{MRT}{2}Q^{-1}\frac{\partial Q}{\partial s_j(q)} \quad (9)$$

with

$$\frac{\partial Q}{\partial s_j(q)} = -2\sum_{m=1}^{M}\sum_{r=1}^{R}\left[W C_{mj}\alpha_{jr} - \left(C_{mj}\alpha_{jr}\right)^2 s_j(q)\right] \quad (10)$$

where

$$W = x_{mr}(q+\tau_{jr}) - \sum_{\substack{n=1 \\ n \neq j}}^{N} C_{mn}\alpha_{nr} s_n(q-\tau_{nr}+\tau_{jr}). \quad (11)$$

The term $W$ is important as it deals with the data, which has been time-shifted according to the latency of the component being estimated, $x_{mr}(q+\tau_{jr})$. From this one subtracts off all the other components after they have been appropriately scaled and time-shifted, $C_{mn}\alpha_{nr} s_n(q-\tau_{nr}+\tau_{jr})$. The derivative of the log probability is zero when the scaled estimated waveshape equals $W$. Thus one can obtain an expression for the optimal waveshape of the $j^{th}$ source in terms of the other sources

$$\hat{s}_j(q) = \frac{\sum_{m=1}^{M}\sum_{r=1}^{R} W C_{mj}\alpha_{jr}}{\sum_{m=1}^{M}\sum_{r=1}^{R}\left(C_{mj}\alpha_{jr}\right)^2}. \quad (12)$$

Similarly for the source amplitudes, one obtains

$$\hat{\alpha}_{jp} = \frac{\sum_{m=1}^{M}\sum_{t=1}^{T}[U V]}{\sum_{m=1}^{M}\sum_{t=1}^{T} V^2} \quad (13)$$

where

$$U = \left(x_{mp}(t) - \sum_{\substack{n=1 \\ n \neq j}}^{N} C_{mn}\alpha_{np} s_n(t-\tau_{np})\right) \quad (14)$$

and

$$V = C_{mj} s_j(t-\tau_{jp}), \quad (15)$$

such that the solution is given by the projection of the detector-scaled component $C_{mj} s_j(t-\tau_{jp})$ onto the data after removing the other scaled and time-shifted components. This is related to matching filter solutions. The optimal source-detector coupling coefficients are found similarly

$$\hat{C}_{ij} = \frac{\sum_{r=1}^{S}\sum_{t=1}^{T}[X\,Y]}{\sum_{r=1}^{S}\sum_{t=1}^{T}Y^2} \quad (16)$$

where

$$X = \left(x_{ir}(t) - \sum_{\substack{n=1\\n\neq j}}^{N} C_{in}\,\alpha_{nr}\,s_n(t-\tau_{nr})\right) \quad (17)$$

and

$$Y = \alpha_{jr}\,s_j(t-\tau_{jr}). \quad (18)$$

Estimating the latency parameter using the approach taken for the other parameters leads to a complex solution as the latency appears implicitly as the argument of the waveshape function. Rather we examine the necessary conditions for maximizing the quadratic form $Q$. Expanding the square in (6), one can see that as the latency $\tau_{jp}$ is varied, only the cross-terms corresponding to the $j^{th}$ source change as long as the source waveshapes are zero outside of a closed time interval. The optimal estimate of the latency $\hat{\tau}_{jp}$ can be found by maximizing

$$Z = \quad (19)$$
$$\sum_{m=1}^{M}\sum_{t=1}^{T}\left[C_{mj}\,\alpha_{jp}\,s_j(t-\tau_{jp})\left(x_{mp}(t) - \sum_{\substack{n=1\\n\neq j}}^{N} C_{mn}\,\alpha_{np}\,s_n(t-\tau_{np})\right)\right],$$

which is the cross-correlation between the estimated source and the data after the contributions from the other sources have been subtracted off. This is then averaged over all the detectors. In practice, as a discrete model is being used for the source waveshapes $\mathbf{s}(t)$, we utilize a discrete set of latencies with resolution equal to the sampling rate.

Iterating equations (12), (13), (16) and (19) over all sources and trials completes the algorithm.

## 4. RESULTS

To demonstrate this approach, we apply the technique to intra-cortically recorded local field potentials from macaque striate cortex during a visuomotor GO - NOGO pattern recognition task [14]. We consider data recorded from only a single channel and show that sufficient information exists to infer multiple component waveforms. The algorithm is easily modified to consider single channel recordings by setting the number of channels $M=1$ and the source-detector coupling coefficients $C_{1n}=1$. The data utilized consists of the ensemble of 222 trials recorded during the GO response.

To demonstrate the degree of the trial-to-trial variability that exists in these data sets, Figure 1 below shows three examples of recorded single trials (noisy waveforms). We calculate the average event-related potential (AERP) by taking the ensemble average of the waveforms in the 222 trials and overlay this waveform on the single-trial waveforms after amplitude scaling and latency shifting according to Eqns. 13 and 19 where the AERP is used as the sole source waveform.

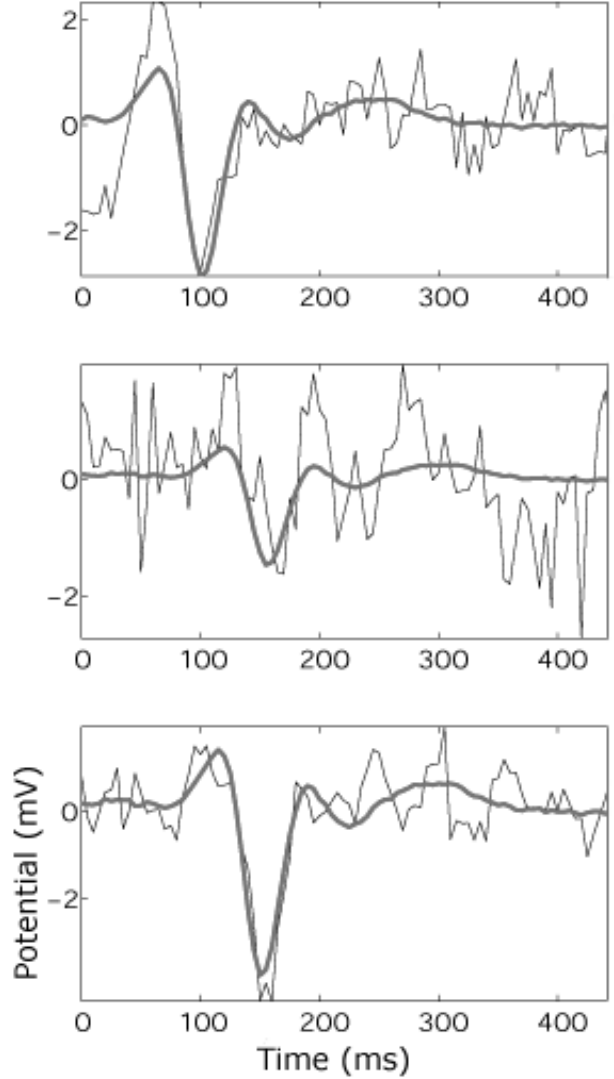

**Fig. 1.** Examples of three single-trial recordings from a striate channel with the AERP overlaid to demonstrate the trial-to-trial variability in the evoked responses.

Examination of the shape of the AERP suggests the contribution of multiple neural sources. In this example we set the number of components to be identified to three, and utilize the local shapes of the AERP around the three extrema as initial guesses for the component waveforms in the MAP estimation. The algorithm will utilize the fact that the different components will exhibit differential variability to identify them. Figure 2 shows the resulting evoked component waveforms. As expected the three components exhibit different variances in their single-trial amplitudes, $\sigma_\alpha^2 = \{0.05, 1.0, 0.14\}$ and latencies $\sigma_\tau^2 = \{24.0, 123.0, 132.6 \ ms^2\}$ for the first, second and third components respectively. An examination of the residual variance, as described in [7], shows that these more detailed models better account for the event-related variability.

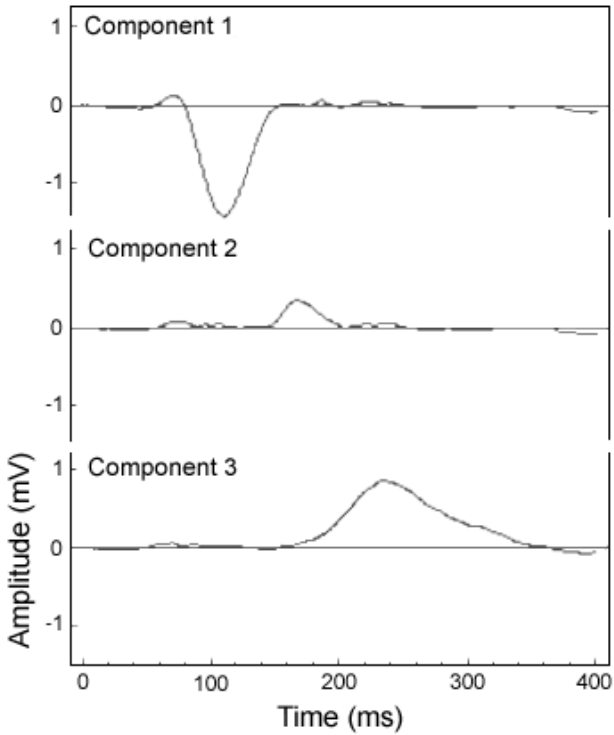

**Fig. 2** Three extracted component waveforms each of which displays unique variability in both amplitude and latency.

It should be noted that although three component waveforms were extracted, one cannot be sure of the number of neural sources responsible for these signals. It is possible that a single neural ensemble generated Component 1 in response to initial feedforward input to the area, followed by Component 3 resulting from subsequent feedback from a higher cortical area. However, it is known that these component waveforms exhibit different trial-to-trial variability. To attempt to identify whether two components are from the same source, one must perform a multiple detector experiment. In that case, the coupling matrix would have two identical columns corresponding to identical coupling between the sources of these components and the array of detectors.

## 5. CONCLUSIONS

In the past, several researchers have utilized maximum likelihood techniques to approach the problem of trial-to-trial variability and source identification [15,16,17,18]. Here with a more detailed model for the evoked responses, we have presented a more general algorithm, which in addition to characterizing responses in single trial data identifies component waveforms based on their differential trial-to-trial variability. The algorithm is derived by approaching the problem as an exercise in model parameter estimation by applying Bayesian inference. There are several advantages to this methodology. First, the strategy is strongly model-based, such that a failure of the algorithm can be traced back to inadequacies of the model to represent the physical situation or to assumptions made in its implementation. Second, any inadequacies of the model, once identified, can be readily remedied given sufficient knowledge about the situation. Third, once the model has been estimated the residual data can be examined to investigate the possibility of additional unimagined phenomena. For instance, this technique seems well suited to the identification of relatively large, low-frequency components. However, it is known that there exist high frequency signals in this data (such as gamma band bursts), but historically their characterization has been difficult. By accurately estimating the contributions from identifiable sources, these effects can be removed to allow researchers to investigate more sensitive signals. Fourth, the Bayesian methodology allows one to incorporate additional prior knowledge into the problem to improve one's inferences.

Comparing the derivation of this algorithm to the Bayesian derivation of Bell and Sejnowski's ICA algorithm [6,9] provides some insights into the relationships between these two techniques. While both algorithms assume that the signal mixing is linear, stationary and instantaneous, this algorithm does not require independence of the sources. This is key in neuroelectrophysiology as the dynamical interactions between neural source generators is a matter of great scientific interest. In addition, this algorithm accommodates noise as well as different numbers of sources and detectors. Finally, by defining the source model to explicitly allow for the potential variability of the source activity in individual trials we make this additional information available to aid in source identification. The algorithm was demonstrated by identifying three components in a set of data recorded from a single channel. As expected, these components exhibited differential variability in both amplitude and latency.

In this algorithm, we chose as a model of the source waveforms a set of discrete points describing the waveform amplitude at regular intervals. This source model is typically used BSS and ICA applications, where the prior probability of these source amplitudes have been given by super- or sub-Gaussian probability densities. It is important to note, for this algorithm and others, that other source models are possible and in many cases desirable. These models could be continuous in nature (especially in the case of continuous latency shifts) such as linear-piecewise or cubic spline models, or could be dynamical in nature such as linear autoregressive moving average (ARMA) models.

Finally, as described in previous works [9,12,13], there are often cases where the experimenter has knowledge about the forward problem, which describes the propagation of the signals to the detectors. In such situations, one can incorporate information about the forward problem into the algorithm along with information about the geometry of the detector array by deriving appropriate prior probabilities for the coupling (or mixing) matrix. In situations where the source locations are of interest, abandoning the coupling matrix in favor of a more detailed model of the source positions and orientations may be more fruitful. By modeling the source locations in addition to the source waveforms, one can easily design an algorithm that simultaneously performs source separation and localization [13].